\documentclass[useAMS,usenatbib]{mn2e}

\usepackage{times}

\usepackage{graphicx}

\title[Fundamental parameters of $\beta$ Vir]
      {The radius and other fundamental parameters of the F9\,V star $\bbeta$~Virginis}

\author[J.R. North et. al.]
       {J.R. North\thanks{E-mail: j.north@physics.usyd.edu.au},
     J.   Davis, 
     J.G. Robertson, 
     T.R. Bedding, 
     H.   Bruntt,
     M.J. Ireland, 
     A.P. Jacob, 
     \newauthor
     S.   Lacour,
     J.W. O'Byrne, 
     S.M. Owens,
     D.   Stello, 
     W.J. Tango and
     P.G. Tuthill\\ 
    Sydney Institute for Astronomy (SIfA), School of Physics, University of Sydney, NSW 2006, Australia}

\newcommand{\half}{{\textstyle\frac{1}{2}}}

\newcommand{\Dnu}{\mbox{$\Delta \nu$}}
\newcommand{\acena}{\mbox{$\alpha$~Cen~A}}
\newcommand{\acenb}{\mbox{$\alpha$~Cen~B}}
\newcommand{\bhyi}{\mbox{$\beta$~Hyi}}
\newcommand{\bvir}{\mbox{$\beta$~Vir}}
\newcommand{\taucet}{\mbox{$\tau$~Cet}}
\newcommand{\muHz}{\mbox{$\mu$Hz}}
\newcommand{\new}[1]{{\bf #1}}
\renewcommand{\new}[1]{{#1}}

\begin{document}

\date{Accepted ; Received ; in original form }

\pagerange{\pageref{firstpage}--\pageref{lastpage}} \pubyear{2008}

\maketitle

\label{firstpage}

\begin{abstract}
We have used the Sydney University Stellar Interferometer (SUSI) to measure
the angular diameter of the F9\,V star $\beta$~Virginis.  After correcting
for limb darkening and combining with the revised Hipparcos parallax, we
derive a radius of $1.703\pm0.022$\,$R_{\sun}$ (1.3~per cent).  We have
also calculated the bolometric flux from published measurements which,
combined with the angular diameter, implies an effective temperature of
$6059 \pm 49$\,K (0.8~per cent).  We also derived the luminosity of \bvir\
to be $L = 3.51\pm0.08\,L_{\sun}$ (2.1~per cent).  Solar-like oscillations
were measured in this star by \citet{CEDAl2005} and using their value for
the large frequency separation yields the mean stellar density with an
uncertainty of about 2~per cent.  Our constraints on the fundamental
parameters of \bvir\ will be important to test theoretical models of this
star and its oscillations.
\end{abstract}


%
%
\section{Introduction}

Accurate measurements of fundamental stellar parameters -- particularly
radius, luminosity and mass -- are essential for critically testing the
current generation of stellar models.  Some of these parameters can be
constrained using well-established techniques such as spectrophotometry,
parallax and optical interferometry.  In the last decade, observing stellar
oscillation frequencies (asteroseismology) has been shown to be a powerful
tool to infer the internal structure of stars (e.g. \citealt{B+G94}).  In
particular, the mean stellar density can be estimated from the large
frequency separation $\Delta\nu$ between consecutive radial overtones.

The links between asteroseismology and interferometry have been
comprehensively reviewed by \citet{CAChD2007}.  Recently, the complementary
roles played by these two techniques in the investigation of solar-type
stars have been described by \citet{CMM2007}, who showed the mass of a
single star can be determined to a precision (1-$\sigma$) of better than 2
per cent by combining a good radius measurement with the density inferred
from the large separation.  Indeed, this precision was almost obtained by
\citet{NDB2007} when they constrained the mass of $\beta$~Hyi to a
precision of 2.8 per cent using results from optical interferometry and
asteroseismology (see also \citealt{KBChD2008}).  Moreover, \citet{NDB2007}
further constrained the fundamental parameters of $\beta$~Hyi with values
for the stellar radius, luminosity, effective temperature and surface
gravity.

The star $\beta$ Virginis (HR 4540, HD 102870, HIP 57757) has spectral type
F9\,V \citep{M+K73} and is slightly metal-rich, with ${\rm [Fe/H]} = 0.14$.
It is bright ($V= 3.60$) and has a well-determined parallax.  Models by
\citet{E+C2006} using the Geneva evolution code, including rotation and
atomic diffusion, indicate that the star has a mass of 1.2--1.3\,M$_{\sun}$
and is close to, or just past, the end of main-sequence core-hydrogen
burning.  Evidence for solar-like oscillations in \bvir\ was reported by
\citet{MLA2004}, in the form of excess power at frequencies around
1.7\,mHz.  Subsequently, \citet{CEDAl2005} confirmed the presence of
oscillations and identified a series of regularly spaced modes with a large
frequency separation of $\Delta\nu=72.1\,\mu$Hz.  

In this paper we present the first interferometric measurement of the
angular diameter of \bvir\ (Section~\ref{sec:diam}).  We also calculate the
bolometric flux of the star using data from the literature
(Section~\ref{sec:flux}).  These, together with the parallax from {\em
Hipparcos} and the density from asteroseismology, allow us to constrain the
fundamental stellar parameters of \bvir\ (effective temperature, radius,
luminosity and mass; Section~\ref{sec:params}).

%
%
\section{Angular Diameter}
\label{sec:diam}

\subsection{Interferometric Observations with SUSI}
\label{sec:obs}

Measurements of the squared visibility~$V^2$ (the normalised
squared-modulus of the complex visibility) were made on a total of 12
nights using the Sydney University Stellar Interferometer (SUSI;
\citealt{DTB99}).  We recorded interference fringes with the red-table
beam-combination system using a filter with (nominal) centre wavelength
700\,nm and full-width at half-maximum 80\,nm. This beam-combination system
has been described in \citet{DIC2007} along with details of the standard
SUSI observing, data reduction and calibration procedures.

The parameters of the calibrator stars are given in Table~\ref{tab:cal}.
\new{There are no measured angular diameters for these stars, and the
values given here were estimated from $(B-V)_0$ colours by interpolating
measurements of similar stars with the Narrabri Stellar Intensity
Interferometer \citep{HDA74} and the Mark III Optical Interferometer
\citep{MAH2003}.  Corrections for limb-darkening were done using the
results of \citet{DTB2000}.}

\begin{table}
\centering
    \caption{Parameters of calibrator stars.}
    \label{tab:cal}
    \begin{tabular}{@{}cccccc@{}}
    \hline
    HR   & Name     & Spectral & V    & UD Diameter       & Separation\\
     &      &   Type   &      &  (mas)            & from $\beta$ Vir\\
    \hline
    4368 & $\phi$ Leo   & A7IV   & 4.47 & $0.58 \pm 0.08$ &  10\fdg08 \\
    4386 & $\sigma$ Leo & B9V    & 4.04 & $0.42 \pm 0.06$ &   8\fdg52 \\
    4515 & $\xi$ Vir    & A4V    & 4.84 & $0.45 \pm 0.06$ &   6\fdg63 \\
    \hline
    \end{tabular}
\end{table}

\begin{table*}
  \centering
  \begin{minipage}{145mm}
    \caption{Summary of observational data.  The night of the observation is
             given in columns 1 and 2 as a calendar date and a mean MJD.
         The nominal and mean projected baseline in units of metres is
         given in columns 3 and 4, respectively. The weighted-mean squared
         visibility, associated error and number of observations during
         a night are given in the last three columns.}
    \label{tab:obs}
    \begin{tabular}{@{}lccclccc}
    \hline
Date    & MJD   & Nominal & Mean Projected  & Calibrators & ${\bar V^2}$ & ${\bar \sigma}$ & \# $V^2$\\
    &   & Baseline & Baseline  &         &  & &\\
    \hline
2007 February 14 & 54145.70 & 40 & 34.05 & $\sigma$ Leo, $\phi$ Leo, $\xi$ Vir & 0.684 & 0.030 & 3\\
2007 February 15 & 54146.66 & 60 & 51.28 & $\sigma$ Leo, $\phi$ Leo & 0.567 & 0.023 & 8\\
2007 March 09    & 54168.58 &  5 &  4.28 & $\sigma$ Leo, $\phi$ Leo & 1.006 & 0.037 & 3 \\
2007 March 10    & 54169.58 & 60 & 51.35 & $\sigma$ Leo, $\phi$ Leo & 0.524 & 0.029 & 5 \\
2007 March 11    & 54170.62 & 15 & 12.74 & $\sigma$ Leo, $\phi$ Leo & 0.893 & 0.018 & 5 \\
2007 March 13    & 54172.63 &  5 &  4.26 & $\sigma$ Leo, $\phi$ Leo & 0.956 & 0.015 & 7 \\
2007 March 14    & 54173.56 & 40 & 34.47 & $\sigma$ Leo, $\phi$ Leo & 0.706 & 0.015 & 6 \\
2007 March 14    & 54173.68 & 15 & 13.25 & $\sigma$ Leo, $\phi$ Leo & 0.951 & 0.063 & 2 \\
2007 March 15    & 54174.59 & 80 & 68.16 & $\sigma$ Leo, $\phi$ Leo & 0.264 & 0.010 & 8 \\
2007 April 18    & 54108.47 &  5 &  4.31 & $\sigma$ Leo, $\phi$ Leo & 0.998 & 0.014 &10 \\
2007 April 20    & 54110.49 & 20 & 17.04 & $\sigma$ Leo, $\phi$ Leo & 0.957 & 0.032 & 4 \\
2007 April 22    & 54112.49 & 80 & 68.01 & $\sigma$ Leo, $\phi$ Leo & 0.339 & 0.014 & 6 \\
2007 May 27      & 54147.39 & 80 & 68.32 & $\sigma$ Leo, $\phi$ Leo & 0.304 & 0.009 & 10 \\
     \hline
    \end{tabular}
 \end{minipage}
\end{table*}

The journal of observations is given in Table~\ref{tab:obs}.  We obtained a
total of 77 estimations of $V^2$.  It should be noted that, for
consistency, the uncertainties in the angular diameters of the calibrating
stars were included in the $V^2$ uncertainty calculation, even though they
have a negligibly small effect when compared to the measurement
uncertainty.

\subsection{Analysis of Visibilities}
\label{sec:analysis}

In the simplest case, the brightness distribution of a star can be modeled
as a disc of uniform irradiance with angular diameter $\theta_{\rm UD}$. The
theoretical response of a two-aperture interferometer to such a model is
given by
\begin{equation}
\label{eq:udisc_v2}
|V|^2 = \left |\frac{2 J_1(\pi B \theta_{\rm UD} / \lambda)}
            {\pi B \theta_{\rm UD} / \lambda} \right|^2,
\end{equation}
\new{where $B$ is the projected baseline,}
$\lambda$ is the observing wavelength and $J_1$ is a first order Bessel
function. However, real stars are limb-darkened and corrections are needed
to find the `true' angular diameter. These corrections can be found in
\citet*{DTB2000} and are small for stars that have a compact atmosphere.

To account for any systematic effects arising from the slightly differing
spectral types of $\beta$~Vir and the calibrators, the presence of a close
companion, or other causes, an additional parameter, $A$, is included in
the uniform disc model. Equation~(\ref{eq:udisc_v2}) becomes
\begin{equation}
\label{eq:udisc2_v2} |V|^2 = \left |A\frac{2 J_1(\pi B
\theta_{\rm{UD}} / \lambda)}
            {\pi B \theta_{\rm{UD}} / \lambda} \right|^2.
\end{equation}

The interferometer response given in equation~(\ref{eq:udisc2_v2}) is,
strictly speaking, only valid for monochromatic observations.
\citet{T+D2002} have analysed the effect of a wide observing bandpass on
interferometric angular diameters and found it to be insignificant provided
the interferometer's coherent field-of-view is larger than the angular
extent of the source.  The coherent field-of-view of SUSI during observations
was found to be greater than 5.8\,mas, hence greater than the extent of
$\beta$~Vir (see below). Therefore, bandwidth smearing can be considered
negligible.  However, SUSI's effective wavelength when observing an F9
star is approximately $696.4\pm 2.0$\,nm \citep{DIC2007}.  The fit to
equation~(\ref{eq:udisc2_v2}) was completed with this effective wavelength.

An implementation of the Levenberg-Marquardt method was used to fit
equation~(\ref{eq:udisc2_v2}) to all measures of $V^2$ to estimate
$\theta_{\rm UD}$ and $A$.  The diagonal elements of the covariance matrix
(which are calculated as part of the $\chi^2$ minimisation) were used to
derive formal uncertainties in the parameter estimation.  We used Markov
chain Monte Carlo (MCMC) simulations to verify the formal uncertainties
because equation~(\ref{eq:udisc2_v2}) is non-linear and the squared
visibility measurement errors (derived by the SUSI reduction pipeline) may
not strictly conform to a normal distribution. A MCMC simulation,
implemented with a Metropolis-Hastings algorithm, involves a
likelihood-based random walk in parameter space whereby the full marginal
posterior probability density function is estimated for each parameter (see
\citealt{Gre2005} for an introduction to MCMC).  An advantage of this
method is that prior information (and the effect on parameter
uncertainties) can be included in the simulation. In the case of
$\beta$~Vir, the effect of the observing wavelength uncertainty was
included in the simulation by the adoption of an appropriate likelihood
function (see below).

\begin{figure}
\includegraphics[width=\linewidth]{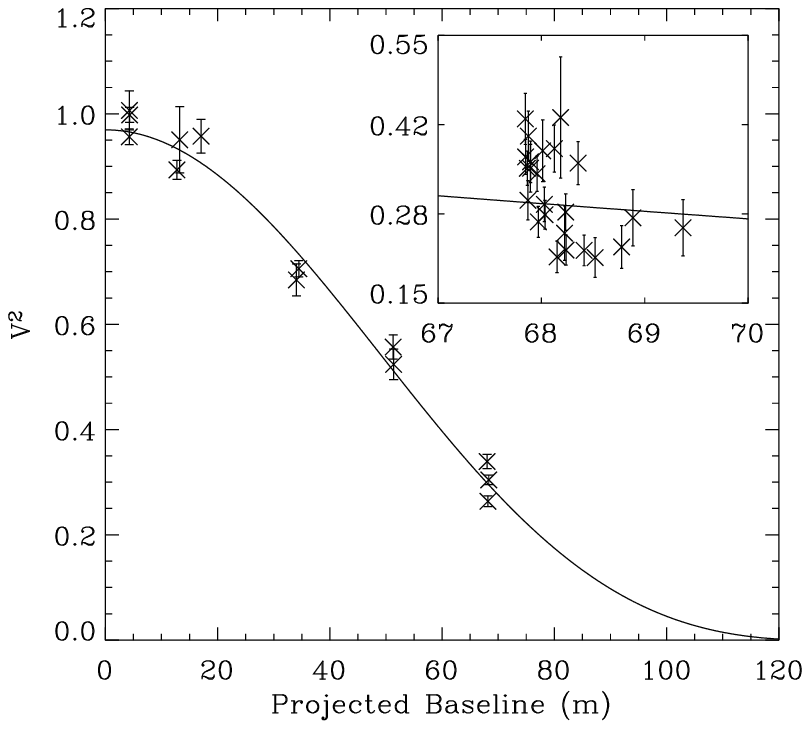}
\caption{\label{fig:vis} Nightly weighted-mean $V^2$ measures with
the fitted uniform disc model overlaid. Inset: all $V^2$ measures at the
(nominal) 80\,m SUSI baseline.}
\end{figure}

The reduced $\chi^2$ of the original fit was 2.08, implying that the
squared-visibility measurement uncertainties were underestimated.  We have
therefore scaled the measurement uncertainties by $\sqrt{2.08}$ to obtain a
reduced $\chi^2$ of unity. The formal uncertainties were verified with a
series of MCMC simulations, each comprising $10^6$ iterations.  For the
effective wavelength we adopted a Gaussian likelihood function with mean
696\,nm and standard deviation 2\,nm.  The probability density functions
produced Gaussian distributions with standard deviation of similar values
to the formal uncertainties.  We therefore obtain (with the associated
$1\sigma$ formal uncertainty) $\theta_{\rm UD} = 1.386\pm0.017$\,mas and $A
= 0.985\pm0.006$.  The data are shown in Fig.~\ref{fig:vis} with the
fitted uniform-disc model overlaid.

The fitted value of $A$ differs slightly from unity, indicating there is
either a close companion or some instrumental effect causing a small
reduction in the $V^2$ intercept.  While $\beta$~Vir does have known
optical companions, these stars are too distant and too faint to affect the
SUSI observations. Furthermore, $\beta$~Vir appears in planet search
programs (see for example \citealt{WEC2006}) and there is only a very
limited parameter space where a further companion could reside without
having already been detected. We also cannot explain the value of $A$ in
terms of a systematic reduction of $V^2$ due to the slightly differing
spectral types of target and calibrator stars.  We believe it is more
likely that the observations were affected by a small unknown systematic
error or a binary companion of unusual characteristics, than that they were
subject to unlucky large random errors.  Therefore, we retain the original
uniform disc angular diameter, $\theta_{\rm UD} = 1.386\pm0.017$\,mas,
fitted with $A$ as a free parameter.

The correction for the effects of limb-darkening was determined to be
$\theta_{\rm LD} = 1.046\theta_{\rm UD}$, using \citet{DTB2000} and the
following parameters: $T_{\rm eff} = 6059$\,K, $\log g = 4.12$ and ${\rm
[Fe/H]} = 0.14$.  With the exception of [Fe/H], these values are the result
of an iterative procedure whereby initial values were refined by the
subsequent determination of fundamental parameters in
Section~\ref{sec:params}.  The starting values for $T_{\rm eff}$ and $\log
g$, and the fixed value for [Fe/H], were the mean values from those found
in the literature, and are given in Table~\ref{tab:lit}. The refinement of
$T_{\rm eff}$ and $\log g$ only caused a small change in the fifth
significant figure of the limb-darkening correction, for which we estimate
an uncertainty 0.002. Therefore the limb-darkened diameter is $\theta_{\rm
LD} = 1.450\pm0.018$\,mas and carries the caveat that we are assuming that
the Kurucz models, upon which \citet{DTB2000} based their work, are
accurate for this star.

\begin{table}
    \caption{Literature values for parameters of $\beta$~Vir.}
    \label{tab:lit}
    \begin{tabular}{@{}llcll@{}}
    \hline
    Source  & $T_{\rm eff}$ (K) & Method$^{\#}$ & $\log g$ & [Fe/H] \\
    \hline
    1   & $6150\pm100$ & a & $4.2\pm0.1$ & 0.10 \\
    2   & $6190\pm80$  & b & $4.20$      & 0.13 \\
    3   & $6176$       & b & $4.14$      & 0.13 \\
    4   & $6127\pm55$  & c &   -         &   -  \\
    5   & $6124\pm31$  & c &   -         &   -  \\
    6   & $6124^\dagger$     &   & $4.24^\dagger$     & $0.19^\dagger$\\
    7   & $6068\pm70$  & b & $4.09\pm0.1$ &$0.13\pm0.1$\\
    8   & $6140$       & d & $4.09\pm0.08$& 0.15 \\
    9   & $6055\pm48$  & e &   -         &   -  \\
    10  & $6124$       & f &     -        & $0.13\pm0.10$ \\
    11  & $6076\pm119$ & b & $4.142$      & 0.13 \\
    \hline
    Mean    & 6123  &  & $4.16$       & 0.14\\
    \hline
    \end{tabular}
    \newline
    Source references: (1)~\citet*{TVV86};  (2)~\citet{Bal90}; (3)~\citet{EAG93};
    (4)~\citet{DiB98};   (5)~\citet{B+LG98}; (6)~\citet{MMB2000};
    (7)~\citet{CNB2001}; (8)~\citet{GGH2001};   (9)~\citet{KSB2003};
    (10)~\citet{M+M2004}; (11)~\citet{APBL2004}.

    $^\dagger$: mean of listed values from elsewhere, many identical;

    $^\#$: method used for $T_{\rm{eff}}$ determination: (a)~Fit to
    H$\gamma$ profile; (b)~From $b$ and ($b-y$) calibration; (c)~IRFM
    method; (d)~Fits of model spectra to spectra and $uby$; (e)~From line
    depth ratios; (f)~From iron line excitation and ionization
    equilibrium.
    \end{table}

%
%
\section{Bolometric Flux}
\label{sec:flux}

We have determined the bolometric flux, $f$, based on flux-calibrated
photometry and spectrophotometry from the literature in combination with
MARCS model stellar atmospheres \citep{GEE2003}.  It has been determined by
summing the integrated fluxes over four spectral ranges, namely
$<0.33\,\mu$m, 0.33--0.86\,$\mu$m, 0.86--2.2\,$\mu$m and 2.2--20\,$\mu$m.
The flux beyond 20\,$\mu$m is $<0.01$~per cent for a 6150\,K black body and will
be similar for the stellar flux distribution and can be ignored.  It is
noted that the ($B-V$) and ($U-B$) colours of $\beta$~Vir are
consistent with those of an unreddened star of its spectral classification.

\citet{Kie87} has published spectrophotometry of $\beta$\,Vir for the
wavelength range 325--865\,nm.  The observations were made at equal
intervals of 1\,nm with a resolution of 1\,nm. The published spectral
energy distribution is averaged over bandpasses 5\,nm wide and is tabulated
every 5\,nm, \new{based on the spectrophotometric calibration of Vega by
\citet{Hay85}.}  This has been converted to a flux distribution using a
value for the flux from Vega at 550.0\,nm of
3.56$\times$10$^{-11}$\,Wm$^{-2}$nm$^{-1}$ \citep{Meg95}.  In the case of
$\delta$~CMa \citep{DBI2007}, Kiehling's spectrophotometry was compared
with that of \citet{D+W74} and the calibrated flux distributions were found
to be in excellent agreement, with an RMS difference computed from the
wavelengths in common of $<$ 1.1~per cent and with no systematic
differences over the wavelength range in common (330--808\,nm). The flux
distribution in the MILES library of empirical spectra \citep{SBPJV2006}
for $\beta$~Vir has also been considered and has been flux-calibrated in
the same way as the Kiehling data.  The wavelength coverage of the MILES
flux distribution is 355--740\,nm, less than the Kiehling range of
325--865\,nm.  It is tabulated at 0.9\,nm intervals with a resolution of
0.23\,nm, compared with the data by Kiehling, which were averaged and
tabulated over 5\,nm intervals. The two distributions are in excellent
agreement except for two apparently discrepant points in the Kiehling
distribution at 760\,nm and 765\,nm.  The two flux distributions were
integrated for the common wavelength range of 355--740\,nm, omitting the
discrepant points, and the integrated fluxes agree to within 1~per cent.
The Kiehling flux distribution covers a greater wavelength range and
extends to the ultraviolet data at the short wavelength end and, for these
reasons, it has been used to determine the integrated visual flux.

The uncertainty in the integrated flux has been estimated by combining the
uncertainty in the \citet{Meg95} flux calibration of 0.7~per cent, the
uncertainty in the Vega calibration by \citet{Hay85} of 1.5~per cent and
the uncertainty in the \new{relative} flux distribution of \citet{Kie87},
which is estimated to be $\sim$1.5~per cent.  This latter figure is based
on the previous experience with $\delta$~CMa and the good agreement between
the Kiehling and MILES flux distributions for $\beta$~Vir.  The resultant
uncertainty is $\pm$2.2~per cent and the integrated Kiehling flux for the
wavelength range 0.33--0.86\,$\mu$m is $(5.73 \pm 0.13)\times
10^{-10}\,{\rm Wm}^{-2}$.

There are four flux values for the ultraviolet from the {\em TD1} satellite
\citep{TNJ78} at 156.5, 196.5, 236.5 and 274.0\,nm and these have been
downloaded from SIMBAD. They have been plotted with the \citet{Kie87}
calibrated fluxes at 325 and 330\,nm and a smooth curve drawn through the
six flux points. The flux shortward of 330\,nm is only a small fraction of
the total flux ($<$4~per cent), justifying this simple approach. The area
under the curve shortward of 330\,nm has been integrated to give the
ultraviolet flux equal to $(3.5 \pm 0.2) \times 10^{-11}\,{\rm Wm}^{-2}$
where the uncertainty is conservatively based on the published
uncertainties in the ultraviolet fluxes.

\begin{figure}
\begin{center}
\includegraphics[scale=0.55]{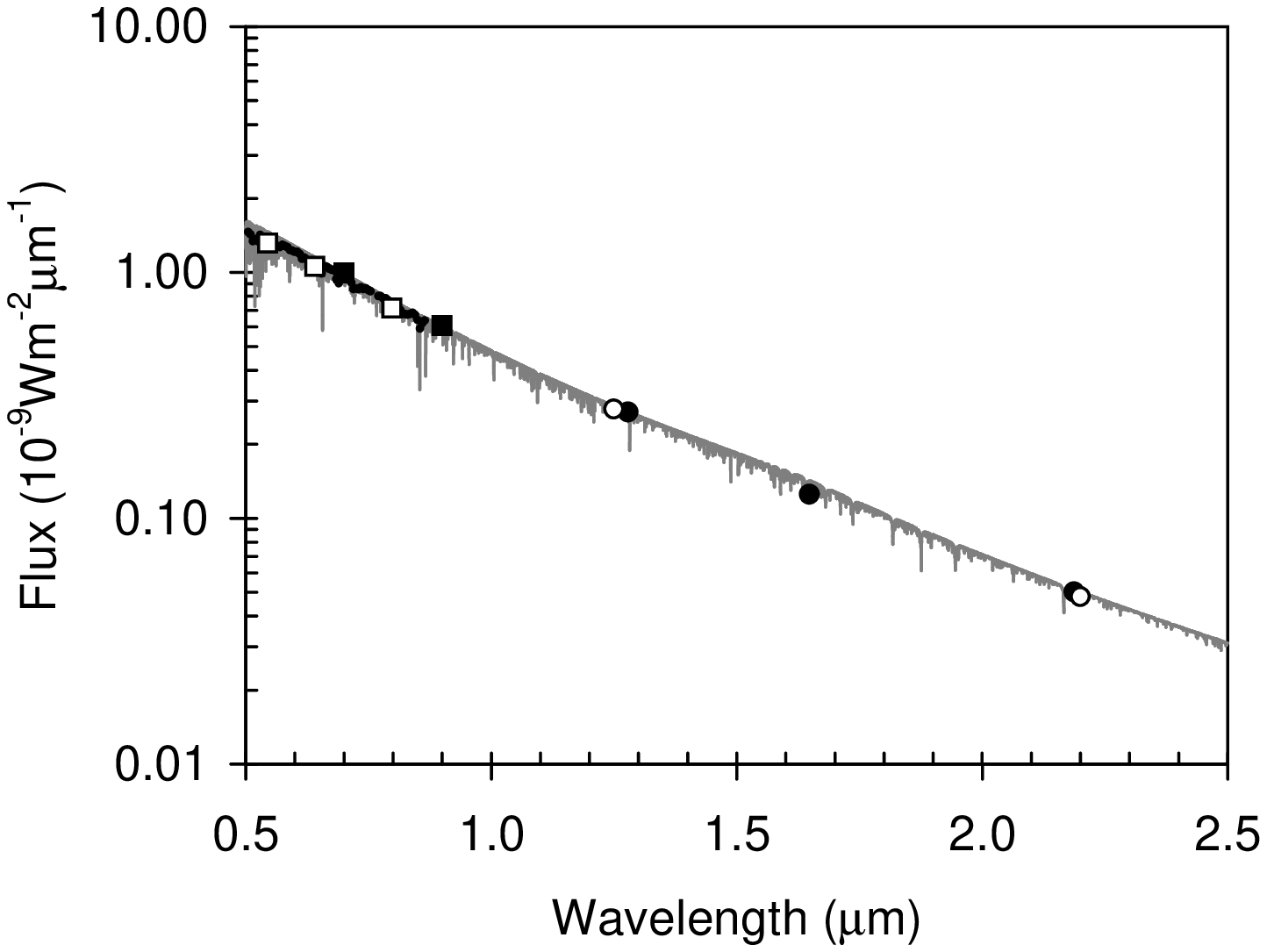}
\caption{The flux distribution for $\beta$~Vir for the wavelength range
  0.5--2.5\,$\mu$m, with data from Table~\ref{tab:phot}.  The symbols have
  been plotted oversize for clarity.  Key: small filled circles -
  \citet{Kie87}; open squares - \citet{Cou80} $VRI$ photometry; filled
  squares - \citet{JIM66} $RI$ photometry; open circles - \citet{JIM66}
  $JK$ photometry; filled circles - \citet{AAMR94} $JHK$ photometry; grey
  line - fluxes for the fitted MARCS model atmosphere averaged over
  0.025~per cent wavelength intervals. }
\label{fig:irflux}
\end{center}
\end{figure}

There are few infrared measurements for $\beta$~Vir and it was necessary to
interpolate and extrapolate them with the aid of a MARCS model stellar
atmosphere \citep{GEE2003}.  The calibrated observational photometric
fluxes are listed in Table~\ref{tab:phot}, together with their sources and
the references used for their calibration, and the fluxes are shown in
Fig.~\ref{fig:irflux}.  Initially the flux distributions of MARCS model
atmospheres for 6000\,K and 6250\,K, both for $\log{g}=4.0$, were scaled to
fit the observational data in the wavelength range 0.6--2.2\,$\mu$m.  The
observational data included the Kiehling fluxes from 0.6--0.865\,$\mu$m and
the fluxes for the $RIJHK$ photometric bands listed in
Table~\ref{tab:phot}.  The 6000\,K model gave a good fit but the 6250\,K
model clearly did not.  The scaled fluxes for the 6000\,K model were
integrated in two ranges, 0.86--2.2\,$\mu$m and 2.2--20\,$\mu$m.  The
resulting integrated fluxes were added to the fluxes for $<$0.33\,$\mu$m
and 0.33--0.86\,$\mu$m determined above and, in combination with the
limb-darkened angular diameter, gave an effective temperature of
$\sim$6050\,K.

\begin{table}
    \caption{Calibrated photometric IR fluxes for $\beta$~Vir
}
    \label{tab:phot}
    \begin{tabular}{@{}ccccc@{}}
    \hline
    Band  & $\lambda_{\rm{eff}}$  & Flux & Source & Calibration \\
          & ($\mu$m) & ($10^{-12}{\rm Wm}^{-2}\mu{\rm m}^{-1}$) & & \\
    \hline
    $R$ & 0.641 & 105.8 & 1 & a \\
    $I$ & 0.798 & 71.6  & 1 & a \\
    $R$ & 0.70  & 99.0  & 2 & b \\
    $I$ & 0.90  & 61.0  & 2 & b \\
    $J$ & 1.25  & 28.0  & 2 & c \\
    $K$ & 2.20  & 4.8   & 2 & c \\
    $J$ & 1.2790 & 27.05 & 3 & d \\
    $H$ & 1.6483 & 12.54 & 3 & d \\
    $K$ & 2.1869 & 5.01  & 3 & d \\
    12$\mu$m & 12 & 0.0104 & 4 & e \\
    \hline
    \end{tabular}
    \newline
Source references: (1)~\citet{Cou80}; (2)~\citet{JIM66};
 (3)~\citet{AAMR94}; (4)~\citet{IRAS88}.

Calibration references: (a)~\citet{BCP98}; (b)~\citet{Joh66};
(c)~\citet{Meg95}; (d)~\citet{AAMR94}; (e)~\citet{IRAS88}
reduced by 4.1~per cent \citep{CWC96}.
\end{table}

The fluxes for an effective temperature of 6050\,K were
interpolated from the 6000\,K and 6250\,K models and the fitting
procedure repeated. The models have fluxes tabulated at intervals
of 0.005~per cent in wavelength, which results in very large plot files.
For diagrammatic purposes the fitting was therefore also carried
out with different flux averages---0.025, 0.05 and 1.0 per cent
intervals in wavelength.  No significant difference was found in
either the scaling factor for a fit or in the integrated fluxes
for the different flux averages.  The fit for the interpolated
fluxes for 6050\,K averaged over 0.025~per cent wavelength intervals is
shown in Fig.~\ref{fig:irflux}.
\new{Uncertainties in the integrated fluxes have been derived by combining
the estimated uncertainty in the fit to the observational data with the
uncertainty in the observational data and in particular that in the
Kiehling data involved in the fit.}

The integrated fluxes for the fitted 6050\,K flux distribution are listed
in Table~\ref{tab:bolflux} for the four wavelength bands considered.  The
uncertainties have been derived by combining the estimated uncertainty in
the fit to the observational data with the uncertainty in the observational
data and in particular that in the Kiehling data involved in the fit.

\begin{figure}
\begin{center}
\includegraphics[scale=0.55]{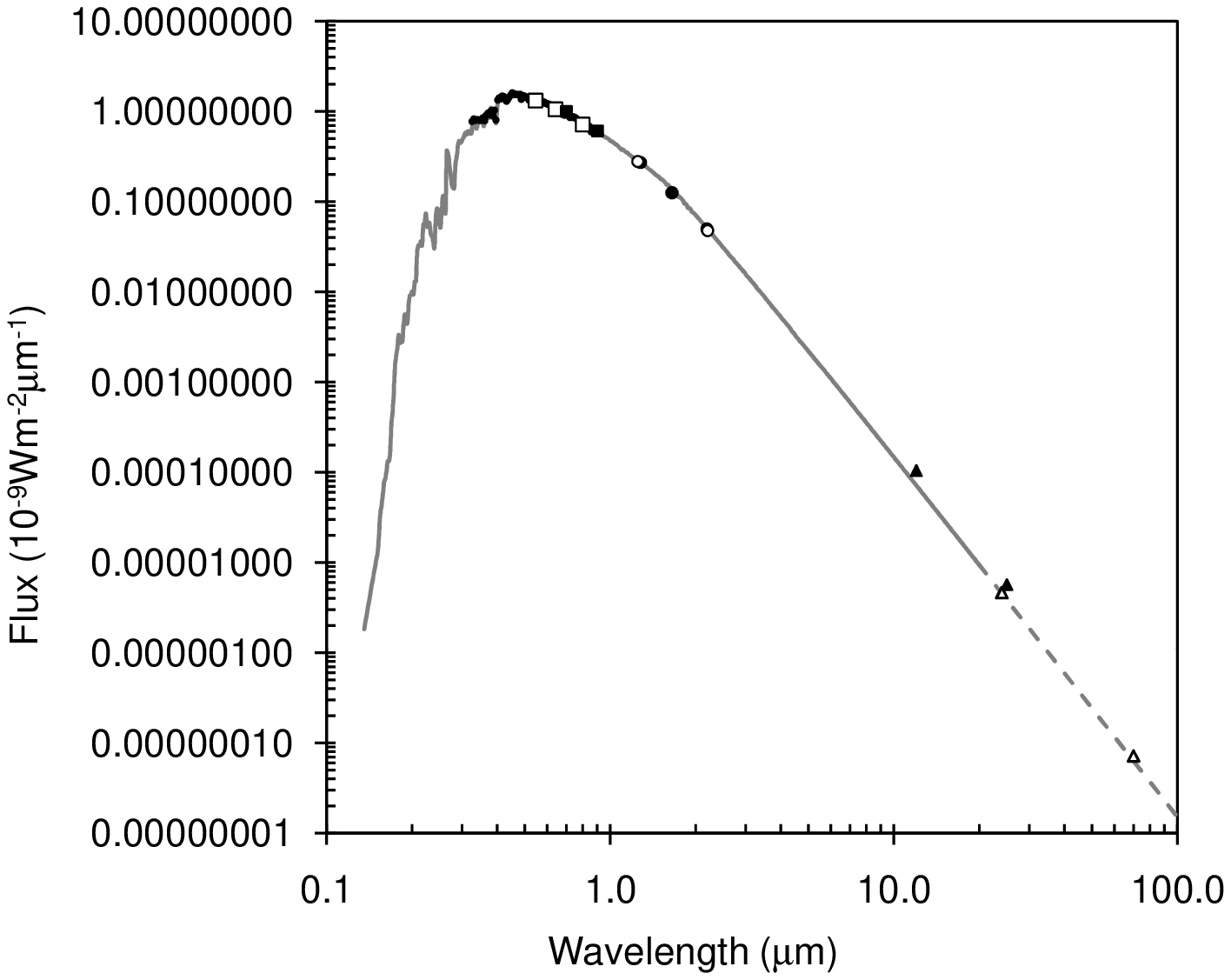}
\caption{The flux distribution for $\beta$~Vir from the ultraviolet to the
mid-infrared.  Symbols have the same meanings as Fig.~\ref{fig:irflux},
with the addition of: filled triangles - {\em IRAS} Point Source Fluxes;
and open triangles - {\em Spitzer} fluxes \citep{TBB2008}.  The dashed grey
line is an extension of the MARCS model beyond 20\,$\mu$m using a slope of
$\lambda^{-4}$.  }
\label{fig:totalflux}
\end{center}
\end{figure}

Figure~\ref{fig:totalflux} shows an assembly of the observational flux data
with the interpolated flux curve for a 6050\,K model atmosphere fitted to
the observational data as shown in Fig.~\ref{fig:irflux}.
\new{Figure~\ref{fig:totalflux} includes mid-infrared photometry from {\em
Spitzer} by \citet{TBB2008} and we note that the data appear to rule out
their suggestion of a possible excess from a debris disk.}  The 12\,$\mu$m
{\em IRAS} Point Source flux does lie above the model curve but drawing a
smooth curve from the flux for the $K$ band through the {\em IRAS} point
and integrating it over the range 2.2--20\,$\mu$m shows that it has
negligible effect on the bolometric flux ($<$0.04~per cent) or effective
temperature ($<$1\,K).

The resulting bolometric flux is $(9.44 \pm 0.20) \times 10^{-10}\,{\rm
Wm}^{-2}$, where the estimated uncertainty takes into account the fact that
the uncertainties in the integrated fluxes for the wavelength ranges
0.33--0.86\,$\mu$m, 0.86--2.2\,$\mu$m and 2.2--20\,$\mu$m are only
partially independent.  Our value is consistent with two previous
determinations, namely $9.408\times10^{-10}$\,Wm$^{-2}$ by \citet{AAMR95}
and $9.59\times10^{-10}$\,Wm$^{-2}$ by \citet{B+LG98}.

\begin{table}
    \caption{Integrated fluxes for $\beta$~Vir in each spectral band
    and the resulting bolometric flux ($f$).
}
    \label{tab:bolflux}
    \begin{tabular}{@{}cc@{}}
    \hline
    Wavelength  & Flux \\
    Range & (10$^{-10}$Wm$^{-2}$) \\
    ($\mu$m) &   \\
    \hline
    0--0.33    & $0.35 \pm 0.02$ \\
    0.33--0.86 & $5.73 \pm 0.13$ \\
    0.86--2.2  & $2.97 \pm 0.09$ \\
    2.2--20    & $0.39 \pm 0.02$ \\
    $>$20      & negligible \\
    \hline
    $f$        & $9.44 \pm 0.20$ \\
    \hline
    \end{tabular}
\end{table}


\begin{table}
    \caption{Physical parameters of $\beta$~Vir.  Estimates for density,
    mass and $\log g$ are given for two different values of $\Dnu$ (see
    text).}
    \label{tab:params}
    \begin{tabular}{@{}lcc@{}}
    \hline
    Parameter & Value  & Uncertainty (per cent) \\
    \hline
    $\theta_{\rm LD}$ (mas) & $1.450\pm0.018$    & 1.2 \\
    $f$ ($10^{-10}$\,Wm$^{-2}$)&$9.44\pm0.20$  & 2.1    \\
    $\pi_{\rm p}$ (mas)     &  $91.50\pm0.22$\rlap{$^a$}   & 0.24 \\
    \hline
    $R$ (R$_{\sun}$) 		& $1.703\pm 0.022$  & 1.3  \\
    $T_{\rm eff}$ (K) 		& $6059 \pm 49$     & 0.8  \\
    $L$ (L$_{\sun}$) 		& $3.51 \pm 0.08$   & 2.1  \\
    \hline
    \Dnu\ (\muHz)               & $72.07 \pm 0.10$   & 0.14 \\
    $\bar{\rho}$ (g\,cm$^{-3}$) & $0.4028\pm 0.0081$ & 2.0  \\
    $M$ (M$_{\sun}$)           & $1.413 \pm 0.061 $ & 4.3 \\
    $\log (g/{\rm cm\,s}^{-2})$ & $4.125 \pm 0.010 $ & 2.4\rlap{ (in $g$)} \\
    \hline
    \Dnu\ (\muHz)               & $70.5$   &  \\
    $\bar{\rho}$ (g\,cm$^{-3}$) & $0.3851\pm 0.0077$ & 2.0  \\
    $M$ (M$_{\sun}$)           & $1.351 \pm 0.058 $ & 4.3 \\
    $\log (g/{\rm cm\,s}^{-2})$ & $4.106 \pm 0.010 $ & 2.4\rlap{ (in $g$)} \\
    \hline
    \end{tabular}
    \newline
    $^{a}$ \citet{vanLee2007}
\end{table}

%
%
\section{Stellar Parameters}
\label{sec:params}

The observable quantities, limb-darkened angular diameter
($\theta_{\mathrm{LD}}$), bolometric flux ($f$) received from the star,
parallax ($\pi_{\rm p}$) and large separation ($\Delta\nu$) combine to
produce experimental constraints on the stellar radius ($R$), effective
temperature ($T_{\rm eff}$), luminosity ($L$), mean density ($\bar{\rho}$),
mass ($M$) and surface gravity ($\log g$).  Values for these observable
quantities are given in Table~\ref{tab:params}, along with the stellar
parameters that are calculated in the remainder of this Section.

\subsection{Radius}

The stellar radius can be determined using our limb-darkened angular
diameter and the {\em Hipparcos} parallax,
\begin{equation}
R = \theta_{\mathrm{LD}} \frac{C}{2\pi_{\rm p}},
\end{equation}
where $C$ is the conversion from parsecs to metres, $\theta_{\mathrm{LD}}$
is in radians and $\pi_{\rm p}$ is in arcsec.  Note that the revised {\em
Hipparcos} parallax of $91.50\pm0.22$\,mas \citep{vanLee2007} has a
substantially smaller uncertainty than the original value of $91.74 \pm
0.77$ \citep{PLK97}.  We obtain a stellar radius for $\beta$~Vir of
$1.703\pm0.022$\,$R_{\sun}$.  This is in excellent agreement with
$1.706\pm0.037$\,R$_{\sun}$ produced from \new{surface brightness}
relations and the {\em Hipparcos} parallax by \citet{TBK2006}, but is more
precise and is based on a direct measurement.

\subsection{Effective Temperature}

The combination of the stellar bolometric flux and angular
diameter yields an empirical effective temperature:
\begin{equation}
\label{eq:teff}
T_{\rm eff} = \left(
\frac{4f}{\sigma\theta_{\mathrm{LD}}^2}\right)^{1/4},
\end{equation}
where $\sigma$ is the Stefan-Boltzmann constant.  Combining the new value
for the bolometric flux (Section~\ref{sec:flux}) with the limb-darkened
angular diameter (Section~\ref{sec:diam}) gives the effective temperature
for $\beta$~Vir as $6059 \pm 49$\,K.

The new value for the effective temperature can be compared with the
previous estimates listed in Table~\ref{tab:lit}.  There is excellent
agreement with the line-depth ratio determination by \citet{KSB2003} and
the determinations by \citet{CNB2001} and \citet{APBL2004} from the
Str\"omgren photometry using the calibrations by \citet{AAMR96} and
\citet{AP+L99}. The methods are completely independent, with only the new
value presented here being based on a direct measurement of the angular
diameter of the star.

The values for the bolometric flux of $9.408 \times 10^{-10}\,{\rm
Wm}^{-2}$ by \citet{AAMR95} and $9.59 \times 10^{-10}\,{\rm Wm}^{-2}$ by
\citet{B+LG98} led those authors to effective temperatures of 6088\,K and
6124\,K respectively.  However, if their bolometric fluxes are combined
with our value for the limb-darkened angular diameter they give effective
temperatures of 6054\,K and 6083\,K respectively, both of which are
consistent with the value for the effective temperature presented here.

\subsection{Luminosity}

The luminosity of a star can be calculated via
\begin{equation}
L = 4 \pi f\frac{C^2}{\pi_{\rm p}^2},
\end{equation}
where $C$ is the conversion from parsecs to metres and $\pi_{\mathrm{p}}$
is in arcsec.  Substituting our value for $f$ gives $L =
(1.349\pm0.029)\times10^{26}$\,W.  Following the work of \citet{BPB2001},
we have adopted L$_{\sun} = (3.842 \pm 0.015)\times 10^{26}$\,W and so get
$L = 3.51\pm0.08\,L_{\sun}$.

A recent determination of $\beta$~Vir's luminosity by \citet{E+C2006},
based on applying a bolometric correction to the absolute visual magnitude,
gave $3.51\pm0.10$\,L$_{\sun}$, which agrees with our value.


\subsection{Mean Density from Asteroseismology}
\label{sec:dens}

\citet{CEDAl2005} collected high-precision velocity measurements of \bvir\
over eleven nights with the CORALIE spectrograph.  The Fourier spectrum
showed a clear power excess in a broad frequency range centred at 1.5\,mHz,
confirming the presence of solar-like oscillations first reported by
\citet{MLA2004}.  \citeauthor{CEDAl2005} found that the autocorrelation of
their power spectrum showed peaks at 70.5, 72 and 74\,\muHz, and they
identified the second of these as the most likely value for the large
frequency separation.  After extracting oscillation frequencies and fitting
to the asymptotic relation \citep{Tas80}, they reported a value for the
large separation of $72.07 \pm 0.10\,\muHz$.

To a good approximation, the mean stellar density can be calculated from
the observed large frequency separation \citep[e.g.][]{Ulr86}:
\begin{equation}
\label{eq:rho}
\frac{\Delta\nu}{\Delta\nu_{\sun}} = \sqrt{
  \frac{\bar{\rho}}{\bar{\rho}_{\sun}}}.
\end{equation}
Using the large separation for \bvir\ reported by \citet{CEDAl2005}, and
adopting solar values of $\bar{\rho}_{\sun} = 1.408$\,g\,cm$^{-3}$ and
$\Delta\nu_{\sun} = 134.8$\,\muHz\ \citep{KBChD2008}, we obtain $\bar{\rho}
= 0.4028$\,g\,cm$^{-3}$.

It is important to note that equation~(\ref{eq:rho}) is an approximation.  In
particular, $\Delta\nu$ in a given star varies systematically with both
frequency and with the angular degree of the modes.  A more sophisticated
approach involves comparing the observed oscillation frequencies with those
calculated from an appropriate stellar model (this was done for $\beta$~Vir
by \citealt{E+C2006}).  Unfortunately, due to the difficulty in modelling
convection in the surface layers of stars, model calculations do not
exactly reproduce observed oscillation frequencies, even for the Sun
\citep{ChDDL88,DPV88,RChDN99,LRD2002}.  This discrepancy increases with
frequency, which means that the large separation is incorrectly predicted
by the model calculations.  For example, as discussed by \citet{KBChD2008},
the best models of the Sun have a large separation that is about 1\,\muHz{}
greater than the observed value.

To address this problem, \citet{KBChD2008} have proposed an empirical
method for correcting the frequencies of stellar models.  They applied this
method to the stars \acena{}, \acenb{} and \bhyi{} and obtained very
accurate estimates of the mean stellar densities (better than 0.5~per cent).  The
same approach was used by \citet{TKB2009} to measure the density of
\taucet\ with similar precision.

We have attempted to apply this method to \bvir, using the oscillation
frequencies published by \citet{CEDAl2005}.  We computed theoretical models
using the Aarhus stellar evolution code (ASTEC, \citealt{ChD2008a}), and
oscillation frequencies using the Aarhus adiabatic oscillation package
(ADIPLS, \citealt{ChD2008b}).  We followed the method described by
\citet{KBChD2008}, which involves fitting both \Dnu\ and the absolute
frequencies of the radial modes (i.e., those having degree $l=0$).  The
method makes use of the fact that the frequency offset between observations
and models should tend to zero with decreasing frequency.  However, we were
not able to achieve a fit with models whose fundamental parameters
(luminosity and radius) agreed with the measured values.  In other words,
the frequencies of the radial modes listed by \citet{CEDAl2005} do not
appear to be consistent with models.  Indeed, for \bvir\ \citet{E+C2006}
remarked on an offset of more than 20\,\muHz\ between observed and
calculated frequencies, which they noted to be substantially larger than
the corresponding offset for the Sun.  Note that our conclusion that the
published frequencies are inconsistent with models is not sensitive to the
input physics of the evolutionary models.  \citet{KBChD2008} demonstrated
that their fitting process works for models published by different authors
using a range of model codes.

Two effects could have introduced a systematic offset to the observed
frequencies.  Firstly, an offset of $\pm 11.6\,\muHz$ could occur because
of the cycle-per-day aliases in the power spectrum, which are strong in
single-site observations such as those obtained for \bvir.  Indeed, many of
the frequencies listed by \citet[][their Table~2]{CEDAl2005} were shifted
by 11.6\,\muHz\ in one direction or the other in order to give a good fit
to the p-mode spectrum.  This includes the highest peak in the power
spectrum, from which 11.6\,\muHz\ was subtracted.

A second possibility is that the modes were misidentified, so that modes of
odd and even degree were interchanged.  That is, modes identified as having
degree $l=0$ actually have degree $l=1$, and vice versa.  Such a
misidentification is plausible and, in the context of the method of
\citeauthor{KBChD2008}, would be equivalent to subtracting an offset of
$\half\Dnu \approx 36\,\muHz$ from the frequencies of all modes.

We found that neither of the above corrections was able to give a
consistent result, but we did find a good agreement between models and
observations if we applied both together.  That is, adding 11.6\,\muHz\ and
subtracting $36\,\muHz$ from the frequencies identified as radial modes
gave a sensible result.  However, if this scenario is correct then the
whole mode identification process would be called into question, and it
would be risky to rely on the tabulated frequencies to establish a precise
density.

Until more oscillation data become available, we therefore fall back on the
density that we determined from equation~(\ref{eq:rho}).  Looking at the
results for other stars \citep{KBChD2008,TKB2009}, we estimate that this
relation gives the mean density with an uncertainty of about 2~per cent.
We have therefore adopted this value for the uncertainty in the density, as
shown in Table~\ref{tab:params}.

\subsection{Mass and Evolutionary State}

Combining our radius measurement with the mean density inferred from
asteroseismology, we calculate the mass to be $1.413 \pm
0.061$\,M$_{\sun}$.  This is slightly higher than previous estimates based
on fitting models in the H-R diagram: $1.36\pm0.09$\,M$_{\sun}$
\citep{AP+L99}, $1.32$\,M$_{\sun}$ \citep{CNB2001} and
$1.34\pm0.10$\,M$_{\sun}$ \citep{L+R2004}.  The difference is not
statistically significant, but we should point out that adopting a value of
$\Dnu = 70.5\,\muHz$, which is \new{another of the three peaks} in the
autocorrelation of the observed power spectrum identified by
\citet{CEDAl2005}, gives a mass of $1.35\,M_{\sun}$.  This value, which is
also shown in Table~\ref{tab:params}, is in better agreement with the
location of the star in the H-R diagram.  The models by \citet{E+C2006}
suggest that \bvir\ either has \new{a mass of $\sim$1.3\,M$_{\sun}$} and is
towards the end of its main-sequence lifetime, or else it has mass of
$\sim$1.2\,M$_{\sun}$ and is in the post main-sequence phase.  Assuming
that one of 70.5 or 72.1\,$\mu$Hz is the correct large spacing, then the
post main-sequence model appears to be excluded, indicating that
$\beta$~Vir is still on the main sequence.  \new{Clearly, establishing the
age of the star is also affected by the uncertainy in the mass.}  A more
precise determination of its mass and evolutionary state will require
further observations of its oscillations, preferably from multiple sites.

Substituting the stellar mass with the mean density and volume, the
standard surface gravity relation becomes
\begin{equation}
g = \frac{4}{3}G\pi\bar{\rho}R,
\end{equation}
where $G$ is the universal constant of gravitation.  In
Table~\ref{tab:params} we show $\log g$ for the two values of $\Dnu$.  Both
determinations are consistent with the published values (see
Table~\ref{tab:lit}), which is not surprising given that spectroscopic
determinations of $\log g$ are not very precise.

\section{Conclusion}

We have presented the first angular diameter measurement of the F9\,V star
$\beta$~Vir. In combination with the revised values for the bolometric flux
and parallax, this angular diameter has experimentally constrained the
stellar radius and effective temperature. The radius measurement, combined
with the mean density determined from asteroseismology, allows an estimate
of $\beta$~Vir's mass and surface gravity.

The constraints on $R$, $T_{\rm eff}$, $M$, $\log g$ and $L$ that we present
will be invaluable in the future to further test theoretical models
of $\beta$~Vir and its oscillations. As stressed by \citet{B+G94}, for
example, oscillation frequencies are of most importance for testing
evolution theories when the other fundamental stellar properties are
well-constrained.

%
%
\section*{Acknowledgments}

This research has been jointly funded by The University of Sydney and the
Australian Research Council as part of the Sydney University Stellar
Interferometer (SUSI) project.  We wish to thank Brendon Brewer for his
assistance with the theory and practicalities of Markov chain Monte Carlo
Simulations.  APJ and SMO acknowledge the support provided by a University
of Sydney Postgraduate Award and a Denison Postgraduate Award,
respectively.  This research has made use of the SIMBAD database, operated
at CDS, Strasbourg, France.

%
%

\bsp

\label{lastpage}


\begin{thebibliography}{}

\bibitem[\protect\citeauthoryear{Allende~Prieto \& Lambert}{Allende~Prieto \&
  Lambert}{1999}]{AP+L99}
Allende~Prieto C.,  Lambert D.~L.,  1999, A\&A, 352, 555

\bibitem[\protect\citeauthoryear{{Allende Prieto}, {Barklem}, {Lambert} \&
  {Cunha}}{{Allende Prieto} et~al.}{2004}]{APBL2004}
{Allende Prieto} C.,  {Barklem} P.~S.,  {Lambert} D.~L.,    {Cunha} K.,  2004,
  A\&A, 420, 183

\bibitem[\protect\citeauthoryear{{Alonso}, {Arribas} \&
  {Martinez-Roger}}{{Alonso} et~al.}{1994}]{AAMR94}
{Alonso} A.,  {Arribas} S.,    {Martinez-Roger} C.,  1994, A\&A, 107, 365

\bibitem[\protect\citeauthoryear{{Alonso}, {Arribas} \&
  {Martinez-Roger}}{{Alonso} et~al.}{1995}]{AAMR95}
{Alonso} A.,  {Arribas} S.,    {Martinez-Roger} C.,  1995, A\&A, 297, 197

\bibitem[\protect\citeauthoryear{{Alonso}, {Arribas} \&
  {Martinez-Roger}}{{Alonso} et~al.}{1996}]{AAMR96}
{Alonso} A.,  {Arribas} S.,    {Martinez-Roger} C.,  1996, A\&A, 313, 873

\bibitem[\protect\citeauthoryear{{Bahcall}, {Pinsonneault} \& {Basu}}{{Bahcall}
  et~al.}{2001}]{BPB2001}
{Bahcall} J.~N.,  {Pinsonneault} M.~H.,    {Basu} S.,  2001, ApJ, 555, 990

\bibitem[\protect\citeauthoryear{{Balachandran}}{{Balachandran}}{1990}]{Bal90}
{Balachandran} S.,  1990, ApJ, 354, 310

\bibitem[\protect\citeauthoryear{Bessell, {Castelli} \& {Plez}}{Bessell
  et~al.}{1998}]{BCP98}
Bessell M.~S.,  {Castelli} F.,    {Plez} B.,  1998, A\&A, 333, 231

\bibitem[\protect\citeauthoryear{{Blackwell} \& {Lynas-Gray}}{{Blackwell} \&
  {Lynas-Gray}}{1998}]{B+LG98}
{Blackwell} D.~E.,  {Lynas-Gray} A.~E.,  1998, A\&AS, 129, 505

\bibitem[\protect\citeauthoryear{Brown \& Gilliland}{Brown \&
  Gilliland}{1994}]{B+G94}
Brown T.~M.,  Gilliland R.~L.,  1994, ARA\&A, 32, 37

\bibitem[\protect\citeauthoryear{{Carrier}, {Eggenberger}, {D'Alessandro} \&
  {Weber}}{{Carrier} et~al.}{2005}]{CEDAl2005}
{Carrier} F.,  {Eggenberger} P.,  {D'Alessandro} A.,    {Weber} L.,  2005,
  NewA, 10, 315

\bibitem[\protect\citeauthoryear{{Chen}, {Nissen}, {Benoni} \& {Zhao}}{{Chen}
  et~al.}{2001}]{CNB2001}
{Chen} Y.~Q.,  {Nissen} P.~E.,  {Benoni} T.,    {Zhao} G.,  2001, A\&A, 371,
  943

\bibitem[\protect\citeauthoryear{Christensen-Dalsgaard}{Christensen-Dalsgaard}%
{2008a}]{ChD2008a}
Christensen-Dalsgaard J.,  2008a, Ap\&SS, 316, 13

\bibitem[\protect\citeauthoryear{{Christensen-Dalsgaard}}{{Christensen-Dalsgaa%
rd}}{2008b}]{ChD2008b}
{Christensen-Dalsgaard} J.,  2008b, Ap\&SS, 316, 113

\bibitem[\protect\citeauthoryear{{Christensen-Dalsgaard}, {D\"appen} \&
  {Lebreton}}{{Christensen-Dalsgaard} et~al.}{1988}]{ChDDL88}
{Christensen-Dalsgaard} J.,  {D\"appen} W.,    {Lebreton} Y.,  1988, Nat, 336,
  634

\bibitem[\protect\citeauthoryear{{Cohen}, {Witteborn}, {Carbon}, {Davies},
  {Wooden} \& {Bregman}}{{Cohen} et~al.}{1996}]{CWC96}
{Cohen} M.,  {Witteborn} F.~C.,  {Carbon} D.~F.,  {Davies} J.~K.,  {Wooden}
  D.~H.,    {Bregman} J.~D.,  1996, AJ, 112, 2274

\bibitem[\protect\citeauthoryear{{Cousins}}{{Cousins}}{1980}]{Cou80}
{Cousins} A.~W.~J.,  1980, SAAO Circ., 1, 234

\bibitem[\protect\citeauthoryear{{Creevey}, {Monteiro}, {Metcalfe}
  et~al.,}{{Creevey} et~al.}{2007}]{CMM2007}
{Creevey} O.~L.,  {Monteiro} M.~J.~P.~F.~G.,  {Metcalfe} T.~S.,    et~al.,
  2007, ApJ, 659, 616

\bibitem[\protect\citeauthoryear{{Cunha}, {Aerts}, {Christensen-Dalsgaard}
  et~al.,}{{Cunha} et~al.}{2007}]{CAChD2007}
{Cunha} M.~S.,  {Aerts} C.,  {Christensen-Dalsgaard} J.,    et~al., 2007,
  A\&AR, 14, 217

\bibitem[\protect\citeauthoryear{{Davis} \& {Webb}}{{Davis} \&
  {Webb}}{1974}]{D+W74}
{Davis} J.,  {Webb} R.~J.,  1974, MNRAS, 168, 163

\bibitem[\protect\citeauthoryear{{Davis}, {Tango}, {Booth}, {ten Brummelaar},
  {Minard} \& {Owens}}{{Davis} et~al.}{1999}]{DTB99}
{Davis} J.,  {Tango} W.~J.,  {Booth} A.~J.,  {ten Brummelaar} T.~A.,  {Minard}
  R.~A.,    {Owens} S.~M.,  1999, MNRAS, 303, 773

\bibitem[\protect\citeauthoryear{{Davis}, {Tango} \& {Booth}}{{Davis}
  et~al.}{2000}]{DTB2000}
{Davis} J.,  {Tango} W.~J.,    {Booth} A.~J.,  2000, MNRAS, 318, 387

\bibitem[\protect\citeauthoryear{{Davis}, {Ireland}, {Chow}, {Jacob}, {Lucas},
  {North}, {O'Byrne}, {Owens}, {Robertson}, {Seneta}, {Tango} \&
  {Tuthill}}{{Davis} et~al.}{2007a}]{DIC2007}
{Davis} J.,  {Ireland} M.~J.,  {Chow} J.,  {Jacob} A.~P.,  {Lucas} R.~E.,
  {North} J.~R.,  {O'Byrne} J.~W.,  {Owens} S.~M.,  {Robertson} J.~G.,
  {Seneta} E.~B.,  {Tango} W.~J.,    {Tuthill} P.~G.,  2007a, PASA, 24, 138

\bibitem[\protect\citeauthoryear{{Davis}, {Booth}, {Ireland}, {Jacob}, {North},
  {Owens}, {Robertson}, {Tango} \& {Tuthill}}{{Davis} et~al.}{2007b}]{DBI2007}
{Davis} J.,  {Booth} A.~J.,  {Ireland} M.~J.,  {Jacob} A.~P.,  {North} J.~R.,
  {Owens} S.~M.,  {Robertson} J.~G.,  {Tango} W.~J.,    {Tuthill} P.~G.,  2007b,
  PASA, 24, 151

\bibitem[\protect\citeauthoryear{{di Benedetto}}{{di Benedetto}}{1998}]{DiB98}
{di Benedetto} G.~P.,  1998, A\&A, 339, 858

\bibitem[\protect\citeauthoryear{{Dziembowski}, {Patern\'o} \&
  {Ventura}}{{Dziembowski} et~al.}{1988}]{DPV88}
{Dziembowski} W.~A.,  {Patern\'o} L.,    {Ventura} R.,  1988, A\&A, 200, 213

\bibitem[\protect\citeauthoryear{Edvardsson, Andersen, Gustafsson, Lambert,
  Nissen \& Tomkin}{Edvardsson et~al.}{1993}]{EAG93}
Edvardsson B.,  Andersen J.,  Gustafsson B.,  Lambert D.~L.,  Nissen P.~E.,
  Tomkin J.,  1993, A\&A, 275, 101

\bibitem[\protect\citeauthoryear{{Eggenberger} \& {Carrier}}{{Eggenberger} \&
  {Carrier}}{2006}]{E+C2006}
{Eggenberger} P.,  {Carrier} F.,  2006, A\&A, 449, 293

\bibitem[\protect\citeauthoryear{{Gray}, {Graham} \& {Hoyt}}{{Gray}
  et~al.}{2001}]{GGH2001}
{Gray} R.~O.,  {Graham} P.~W.,    {Hoyt} S.~R.,  2001, AJ, 121, 2159

\bibitem[\protect\citeauthoryear{Gregory}{Gregory}{2005}]{Gre2005}
Gregory P.~C.,  2005, Bayesian Logical Data Analysis for the Physical Sciences.
Cambridge University Press

\bibitem[\protect\citeauthoryear{{Gustafsson}, {Edvardsson}, {Eriksson},
  {Mizuno-Wiedner}, {J{\o}rgensen} \& {Plez}}{{Gustafsson}
  et~al.}{2003}]{GEE2003}
{Gustafsson} B.,  {Edvardsson} B.,  {Eriksson} K.,  {Mizuno-Wiedner} M.,
  {J{\o}rgensen} U.~G.,    {Plez} B.,  2003, in {Hubeny} I.,  {Mihalas} D.,
  {Werner} K.,  eds, Stellar Atmosphere Modeling Vol.~288 of ASP Conf. Series,
  {A Grid of Model Atmospheres for Cool Stars}.
p.~331

\bibitem[\protect\citeauthoryear{Hanbury~Brown, Davis \& Allen}{Hanbury~Brown
  et~al.}{1974}]{HDA74}
Hanbury~Brown R.,  Davis J.,    Allen L.~R.,  1974, MNRAS, 167, 121

\bibitem[\protect\citeauthoryear{{Hayes}}{{Hayes}}{1985}]{Hay85}
{Hayes} D.~S.,  1985, in {Hayes} D.~S.,  {Pasinetti} L.~E.,   {Philip}
  A.~G.~D.,  eds, Calibration of Fundamental Stellar Quantities Vol.~111 of IAU
  Symp., {Stellar absolute fluxes and energy distributions from 0.32 to 4.0
  microns}.
p.~225

\bibitem[\protect\citeauthoryear{{IRAS Team}}{{IRAS Team}}{1988}]{IRAS88}
{IRAS Team} 1988, {The IRAS Point Source Catalog, version 2.0}.
NASA RP--1190

\bibitem[\protect\citeauthoryear{{Johnson}}{{Johnson}}{1966}]{Joh66}
{Johnson} H.~L.,  1966, ARA\&A, 4, 193

\bibitem[\protect\citeauthoryear{{Johnson}, {Iriarte}, {Mitchell} \&
  {Wisniewskj}}{{Johnson} et~al.}{1966}]{JIM66}
{Johnson} H.~L.,  {Iriarte} B.,  {Mitchell} R.~I.,    {Wisniewskj} W.~Z.,
  1966, Comm. Lunar and Planetary Lab., 4, 99

\bibitem[\protect\citeauthoryear{{Kiehling}}{{Kiehling}}{1987}]{Kie87}
{Kiehling} R.,  1987, A\&AS, 69, 465

\bibitem[\protect\citeauthoryear{Kjeldsen, Bedding \&
  Christensen-Dalsgaard}{Kjeldsen et~al.}{2008}]{KBChD2008}
Kjeldsen H.,  Bedding T.~R.,    Christensen-Dalsgaard J.,  2008, ApJ, 683, L175

\bibitem[\protect\citeauthoryear{{Kovtyukh}, {Soubiran}, {Belik} \&
  {Gorlova}}{{Kovtyukh} et~al.}{2003}]{KSB2003}
{Kovtyukh} V.~V.,  {Soubiran} C.,  {Belik} S.~I.,    {Gorlova} N.~I.,  2003,
  A\&A, 411, 559

\bibitem[\protect\citeauthoryear{{Lambert} \& {Reddy}}{{Lambert} \&
  {Reddy}}{2004}]{L+R2004}
{Lambert} D.~L.,  {Reddy} B.~E.,  2004, MNRAS, 349, 757

\bibitem[\protect\citeauthoryear{{Li}, {Robinson}, {Demarque}, {Sofia} \&
  {Guenther}}{{Li} et~al.}{2002}]{LRD2002}
{Li} L.~H.,  {Robinson} F.~J.,  {Demarque} P.,  {Sofia} S.,    {Guenther}
  D.~B.,  2002, ApJ, 567, 1192

\bibitem[\protect\citeauthoryear{{Malagnini}, {Morossi}, {Buzzoni} \&
  {Chavez}}{{Malagnini} et~al.}{2000}]{MMB2000}
{Malagnini} M.~L.,  {Morossi} C.,  {Buzzoni} A.,    {Chavez} M.,  2000, PASP,
  112, 1455

\bibitem[\protect\citeauthoryear{Marti{\'c}, {Lebrun}, {Appourchaux} \&
  {Korzennik}}{Marti{\'c} et~al.}{2004}]{MLA2004}
Marti{\'c} M.,  {Lebrun} J.-C.,  {Appourchaux} T.,    {Korzennik} S.~G.,  2004,
  A\&A, 418, 295

\bibitem[\protect\citeauthoryear{{Megessier}}{{Megessier}}{1995}]{Meg95}
{Megessier} C.,  1995, A\&A, 296, 771

\bibitem[\protect\citeauthoryear{{Morel} \& {Micela}}{{Morel} \&
  {Micela}}{2004}]{M+M2004}
{Morel} T.,  {Micela} G.,  2004, A\&A, 423, 677

\bibitem[\protect\citeauthoryear{{Morgan} \& {Keenan}}{{Morgan} \&
  {Keenan}}{1973}]{M+K73}
{Morgan} W.~W.,  {Keenan} P.~C.,  1973, ARA\&A, 11, 29

\bibitem[\protect\citeauthoryear{{Mozurkewich}, {Armstrong}, {Hindsley},
  {Quirrenbach}, {Hummel}, {Hutter}, {Johnston}, {Hajian}, {Elias} II,
  {Buscher} \& {Simon}}{{Mozurkewich} et~al.}{2003}]{MAH2003}
{Mozurkewich} D.,  {Armstrong} J.~T.,  {Hindsley} R.~B.,  {Quirrenbach} A.,
  {Hummel} C.~A.,  {Hutter} D.~J.,  {Johnston} K.~J.,  {Hajian} A.~R.,  {Elias}
  II N.~M.,  {Buscher} D.~F.,    {Simon} R.~S.,  2003, AJ, 126, 2502

\bibitem[\protect\citeauthoryear{North, {Davis}, {Bedding} et~al.,}{North
  et~al.}{2007}]{NDB2007}
North J.~R.,  {Davis} J.,  {Bedding} T.~R.,    et~al., 2007, MNRAS, 380, L83

\bibitem[\protect\citeauthoryear{Perryman, Lindegren, Kovalevsky, H{\o}g,
  Bastian, Bernacca, Creze, Donati, Grenon, Grewing, van Leeuwen, van~der
  Marel, Mignard, Murray, Le~Poole, Schrijver, Turon, Arenou, Froeschle \&
  Petersen}{Perryman et~al.}{1997}]{PLK97}
Perryman M. A.~C.,  Lindegren L.,  Kovalevsky J.,  H{\o}g E.,  Bastian U.,
  Bernacca P.~L.,  Creze M.,  Donati F.,  Grenon M.,  Grewing M.,  van Leeuwen
  F.,  van~der Marel H.,  Mignard F.,  Murray C.~A.,  Le~Poole R.~S.,
  Schrijver H.,  Turon C.,  Arenou F.,  Froeschle M.,    Petersen C.~S.,  1997,
  A\&A, 323, L49

\bibitem[\protect\citeauthoryear{{Rosenthal}, {Christensen-Dalsgaard},
  {Nordlund}, {Stein} \& {Trampedach}}{{Rosenthal} et~al.}{1999}]{RChDN99}
{Rosenthal} C.~S.,  {Christensen-Dalsgaard} J.,  {Nordlund} {\AA}.,  {Stein}
  R.~F.,    {Trampedach} R.,  1999, A\&A, 351, 689

\bibitem[\protect\citeauthoryear{{S{\'a}nchez-Bl{\'a}zquez}, {Peletier},
  {Jim{\'e}nez-Vicente}, {Cardiel}, {Cenarro}, {Falc{\'o}n-Barroso}, {Gorgas},
  {Selam} \& {Vazdekis}}{{S{\'a}nchez-Bl{\'a}zquez} et~al.}{2006}]{SBPJV2006}
{S{\'a}nchez-Bl{\'a}zquez} P.,  {Peletier} R.~F.,  {Jim{\'e}nez-Vicente} J.,
  {Cardiel} N.,  {Cenarro} A.~J.,  {Falc{\'o}n-Barroso} J.,  {Gorgas} J.,
  {Selam} S.,    {Vazdekis} A.,  2006, MNRAS, 371, 703

\bibitem[\protect\citeauthoryear{{Tango} \& {Davis}}{{Tango} \&
  {Davis}}{2002}]{T+D2002}
{Tango} W.~J.,  {Davis} J.,  2002, MNRAS, 333, 642

\bibitem[\protect\citeauthoryear{Tassoul}{Tassoul}{1980}]{Tas80}
Tassoul M.,  1980, ApJS, 43, 469

\bibitem[\protect\citeauthoryear{Teixeira, Kjeldsen, Bedding et~al.,}{Teixeira
  et~al.}{2009}]{TKB2009}
Teixeira T.,  Kjeldsen H.,  Bedding T.~R.,    et~al., 2009, A\&A, in press

\bibitem[\protect\citeauthoryear{Th{\' e}venin, {Vauclair} \& {Vauclair}}{Th{\'
  e}venin et~al.}{1986}]{TVV86}
Th{\' e}venin F.,  {Vauclair} S.,    {Vauclair} G.,  1986, A\&A, 166, 216

\bibitem[\protect\citeauthoryear{{Th{\'e}venin}, {Bigot}, {Kervella}, {Lopez},
  {Pichon} \& {Schmider}}{{Th{\'e}venin} et~al.}{2006}]{TBK2006}
{Th{\'e}venin} F.,  {Bigot} L.,  {Kervella} P.,  {Lopez} B.,  {Pichon} B.,
  {Schmider} F.-X.,  2006, Mem. Soc. Astron. Ital., 77, 411

\bibitem[\protect\citeauthoryear{{Thompson}, {Nandy}, {Jamar}, {Monfils},
  {Houziaux}, {Carnochan} \& {Wilson}}{{Thompson} et~al.}{1978}]{TNJ78}
{Thompson} G.~I.,  {Nandy} K.,  {Jamar} C.,  {Monfils} A.,  {Houziaux} L.,
  {Carnochan} D.~J.,    {Wilson} R.,  1978, {Catalogue of stellar ultraviolet
  fluxes (TD-1)}.
The Science Research Council, UK

\bibitem[\protect\citeauthoryear{{Trilling}, {Bryden}, {Beichman}, {Rieke},
  {Su}, {Stansberry}, {Blaylock}, {Stapelfeldt}, {Beeman} \&
  {Haller}}{{Trilling} et~al.}{2008}]{TBB2008}
{Trilling} D.~E.,  {Bryden} G.,  {Beichman} C.~A.,  {Rieke} G.~H.,  {Su}
  K.~Y.~L.,  {Stansberry} J.~A.,  {Blaylock} M.,  {Stapelfeldt} K.~R.,
  {Beeman} J.~W.,    {Haller} E.~E.,  2008, ApJ, 674, 1086

\bibitem[\protect\citeauthoryear{Ulrich}{Ulrich}{1986}]{Ulr86}
Ulrich R.~K.,  1986, ApJ, 306, L37

\bibitem[\protect\citeauthoryear{van Leeuwen}{van Leeuwen}{2007}]{vanLee2007}
van Leeuwen F.,  2007, {Hipparcos, the New Reduction of the Raw Data}.
Springer: Dordrecht

\bibitem[\protect\citeauthoryear{{Wittenmyer}, {Endl}, {Cochran}, {Hatzes},
  {Walker}, {Yang} \& {Paulson}}{{Wittenmyer} et~al.}{2006}]{WEC2006}
{Wittenmyer} R.~A.,  {Endl} M.,  {Cochran} W.~D.,  {Hatzes} A.~P.,  {Walker}
  G.~A.~H.,  {Yang} S.~L.~S.,    {Paulson} D.~B.,  2006, AJ, 132, 177

\end{thebibliography}
\end{document}